\documentclass[lettersize,journal]{IEEEtran}
\usepackage{amsmath,amsfonts}
\usepackage{algorithmic}
\usepackage{algorithm}
\usepackage{array}
\usepackage[caption=false,font=normalsize,labelfont=sf,textfont=sf]{subfig}
\usepackage{textcomp}
\usepackage{stfloats}
\usepackage{url}
\usepackage{verbatim}
\usepackage{graphicx}
\usepackage{cite}
\usepackage{siunitx}

\usepackage{booktabs}
\usepackage{multirow}
\usepackage{graphicx}

\begin{document}

\title{Deep learning model transfer in forest mapping using multi-source satellite SAR and optical images}

\author{Shaojia Ge, Oleg Antropov*, Tuomas Häme, Ronald E. McRoberts, and Jukka Miettinen
\thanks{Shaojia Ge is with Department of Electronic Engineering, School of Electronic and Optical Engineering, Nanjing University of Science and Technology, Nanjing 210094, China (e-mail: geshaojia@njust.edu.cn),}
\thanks{Oleg Antropov, Tuomas Häme and Jukka Miettinen are with VTT Technical Research Centre of Finland, 00076 Espoo, Finland (e-mail: name.surname@vtt.fi),}
\thanks{Ronald E. McRoberts is with Department of Forest Resources, University of Minnesota, Saint Paul, MN 55108, USA (e-mail: mcrob001@umn.edu),}
\thanks{Correspondence: oleg.antropov@vtt.fi.}}



\maketitle

\begin{abstract}
Deep learning (DL) models are gaining popularity in forest variable prediction using Earth Observation images. However, in practical forest inventories, reference datasets are often represented by plot- or stand-level measurements, while high-quality representative wall-to-wall reference data for end-to-end training of DL models are rarely available. Transfer learning facilitates expansion of the use of deep learning models into areas with sub-optimal training data by allowing pretraining of the model in areas where high-quality teaching data are available. In this study, we perform a "model transfer" (or domain adaptation) of a pretrained DL model into a target area using plot-level measurements and compare performance versus other machine learning models. We use an earlier developed UNet based model (SeUNet) to demonstrate the approach on two distinct taiga sites with varying forest structure and composition. Multisource Earth Observation (EO)  data are represented by a combination of Copernicus Sentinel-1 C-band SAR and Sentinel-2 multispectral images, JAXA ALOS-2 PALSAR-2 SAR mosaic and TanDEM-X bistatic interferometric radar data. The training study site is located in Finnish Lapland, while the target site is located in Southern Finland. By leveraging transfer learning, the prediction of SeUNet achieved root mean squared error (RMSE) of \SI{2.70}{m} and R$^2$ of 0.882, considerably more accurate than traditional benchmark methods. We expect such forest-specific DL model transfer can be suitable also for other forest variables and other EO data sources that are sensitive to forest structure.
\end{abstract}


\section{Introduction}
\IEEEPARstart{F}{orests} cover approximately one-third of Earth's landmass (FAO 2022) and play a key role in mitigating the effects of climate change by reducing the concentration of carbon dioxide in the atmosphere. Forests are fundamental for preserving biodiversity as they are the natural habitat to a myriad of plant and animal species. Several international initiatives, such as the Framework Convention on Climate Change\footnote{https://unfccc.int/} and the Convention on Biological Diversity\footnote{https://www.cbd.int/} by the United Nations, and the new Forest (EU 2021) and Biodiversity (EU 2020) strategies by the European Union, induce increased forest monitoring requirements at national and local levels by increasing the reporting requirements. Interest in voluntary certification schemes and diversification of forest uses (e.g. carbon or other ecosystem services) is also growing among forestry stakeholders, which typically requires improved forest monitoring approaches for verification purposes, further increasing the need for reliable high frequency information on forests. 

The use of Earth Observation (EO) data has become an integral part of forest monitoring due to numerous satellite sensor missions designed for environmental monitoring during the last decades \cite{Herold2019, mcrob2007}. However, as satellite sensors typically cannot directly measure many forest attributes, indirect modelling approaches have been developed that combine EO data with ground based observations to provide a set of predictions presented in the form of a map. In wide-area forest mapping, statistical, physics-based, and machine learning (ML) methods have been used for modeling and prediction purposes \cite{gfoi-2014}. Satellite remotely sensed data combined with in-situ forest measurements can be considered a cost-effective means for producing forest attribute maps and forest estimates on various areal levels, but such predictions and estimates can have considerable uncertainty \cite{mcrob2007}.

EO datasets used for mapping of boreal forest resources include both optical and SAR datasets \cite{gfoi-2014, rodriguez2019}. Optical datasets were widely used historically and are particularly suitable for mapping forest cover and tree species\cite{astola2019}. The SAR signal, especially at lower frequencies, is sensitive to forest biomass and structure \cite{gfoi-2014}. Multitemporal and polarimetric features provide improved accuracies \cite{schmullius2015,antropov2017}. Interferometric SAR datasets have particular sensitivity to vertical structure of forest that makes them very suitable predictors for mapping forest structure and height \cite{olsk2016,kugler2015}. 

Deep learning (DL) methods are widely adopted for various image classification, and semantic segmentation tasks \cite{persello2022,zhu2020,zhu2017}. To date, several fully convolutional and recurrent neural networks have been  demonstrated in forest remote sensing \cite{astola2021,illarionova2021,ge2022improved,ge2022lstm,bolyn2022,wang2020,illarionova2022,lang2019country}. These models often provide improved accuracy in forest classification or predicting forest variables, as well as in forest change mapping. However, training of DL models often requires a fully segmented reference label, such as airborne laser scanner (ALS) based forest attribute maps that are costly and not available over wide areas. 

Instead, reference data from forests are typically available as field sample plots measurements. The sample plots have been traditionally used to derive sample based areal estimates of forest resources, but in recent decades also as in-situ data for training and evaluating remote sensing based models. 
 
Within the DL context, such reference data can be considered \textit{weak labels} \cite{wang2020} and they are not optimal for training a DL model. In addition to the typically insufficient number of sample plots, they lack information on the spatial context. 

Although there is an increasing amount of ALS data available over boreal forests, current airborne campaigns do not provide the spatial and temporal coverage required to meet the needs for modern forest inventories. For example, annual information would be needed to support forest management decisions and monitoring requirements in an operational context. Therefore, the lack of appropriate ALS data greatly reduces the utility of DL models for operational forest monitoring. With transfer learning techniques, it may be possible to fine-tune DL models to the area of interest using data for field sample plots that are generally more widely available in managed forest areas. This approach would open possibilities for both geographic (i.e. from an area that has ALS data to an area that has not) and temporal (i.e. from a year that has ALS data to a year that has not) transfer of DL models, facilitating wider usability of DL-based approaches in operational forest monitoring. 

In this study, we aim to demonstrate the potential of transfer learning in the context of forest monitoring with deep learning models. We pretrain an earlier developed SeUNet deep learning model \cite{ge2022improved} with ALS-based data in a training site situated in Lapland in the Northern part of Finland. Further, we apply the pretrained model in the target site in the Southern part of Finland, with and without fine-tuning the model with field sample plots from the target area. We highlight the effects of the model fine-tuning with field plot data and compare the resulting accuracy to traditional machine learning methods trained in the target area using the same field plots. We also investigate scenarios where reference forest plots in the target site are very scarce or measurements for certain forest strata are completely missing. Additionally, we compare various EO sensors in terms of prediction accuracies.

\section{Materials and Methods}

\begin{figure*}[htb]
\centering
\includegraphics[width=0.9\linewidth]{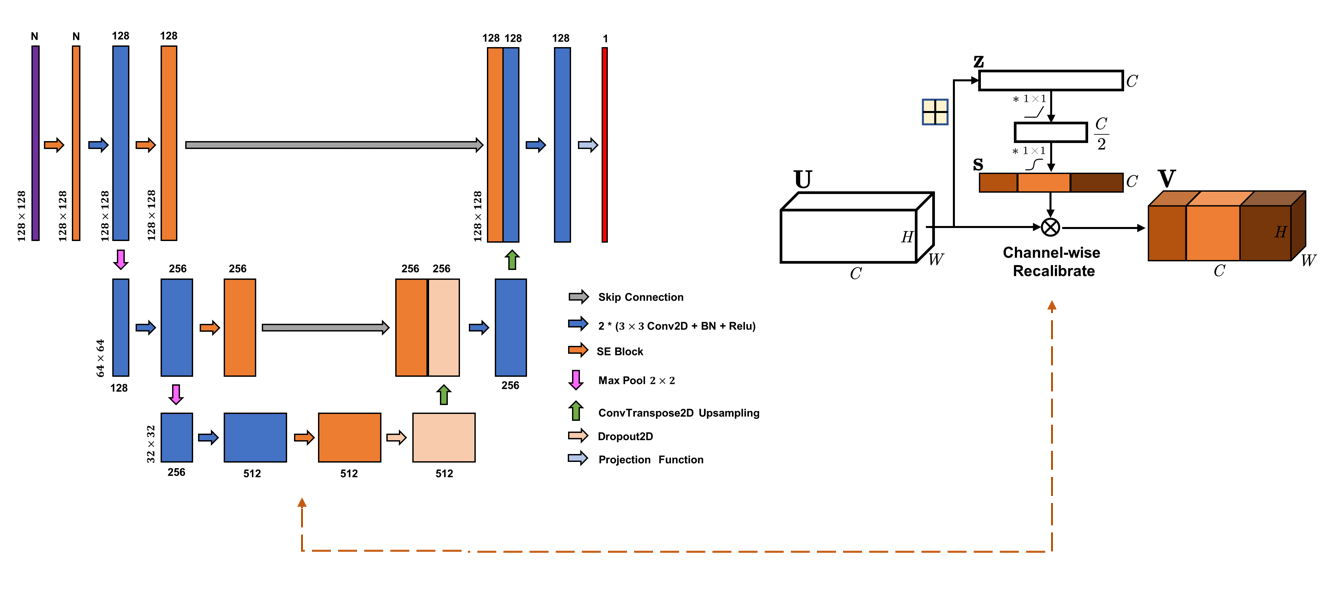}
\caption{SeUNet model for regression task with embedded squeeze-excitation blocks}
\label{fig:seunet}
\end{figure*}

\subsection{UNet model and its derivative/improved models}
One of the main and frequently used architectures of CNNs for semantic segmentation of satellite images, including the forest mask segmentation, is the U-Net architecture.
UNet is a variant of convolutional network originally introduced for biomedical image segmentation \cite{ronneberger2015unet} and is presently often used in various semantic segmentation and regression tasks \cite{scepanovic2021, ge2022improved,illarionova2022}. The basic UNet (also known as Vanilla UNet) uses convolutional network to extract image features. The UNet model consists of an encoder and a decoder, which are connected by skip connections. The encoder is responsible for feature extraction and the decoder is used to restore the feature map to its original size. The model is symmetrical in structure and has a double-convolution structure at its core, which is made up of a 2-D convolution, batch normalization, and ReLU activation. This structure allows UNet to extract deeper features of the input data and maintain a good fusion ability at all levels, while keeping the feature map size unchanged. The overall architecture of UNet makes it well-suited for pixel-level classification and regression tasks.

Here, we use an improved version called SeUNet, suitable for producing spatially explicit pixel-level forest inventory using EO data \cite{ge2022improved} when trained with fully-segmented (spatially explicit) image patches, such as ALS-based forest inventory data. Within SeUNet, a  Squeeze-and-Excitation attention module is used to recalibrate the multi-source features using channel self-attention to improve the accuracy of predictions with limited reference data. SeUNet was superior to basic UNet model and was shown to be particularly effective in boreal forest height mapping using Sentinel-1 time series and Sentinel-2 datasets \cite{ge2022improved}. The structure of the model is shown in Figure \ref{fig:seunet}.

\subsection{How transfer learning is organized}
The concept of transfer learning in deep learning involves using a pre-trained model on a large dataset as a starting point for training on a new, smaller dataset \cite{zhuang2020comprehensive,huang2017transfer,wurm2019semantic}. 
Since both study areas (initial and target sites) are represented by boreal forests, we can safely assume their latent representations strongly overlap in EO feature space, although the specific forest characteristics may be different (considerably more sparse forest in Lapland). This means the prior knowledge learned from the source site can mitigate the negative influence brought by limited reference, e.g. plot-level forest reference that limits inferring spatial context (neighbourhood features). 
Our hypothesis is that by leveraging the knowledge gained from the pre-training on a spatially explicit dataset, we can achieve better results compared to the end-to-end training from scratch using conventional statistical and machine learning approaches with limited reference forest plot data collected at the new target site.

\begin{figure*}[htb]
\centering
\includegraphics[width=0.9\linewidth]{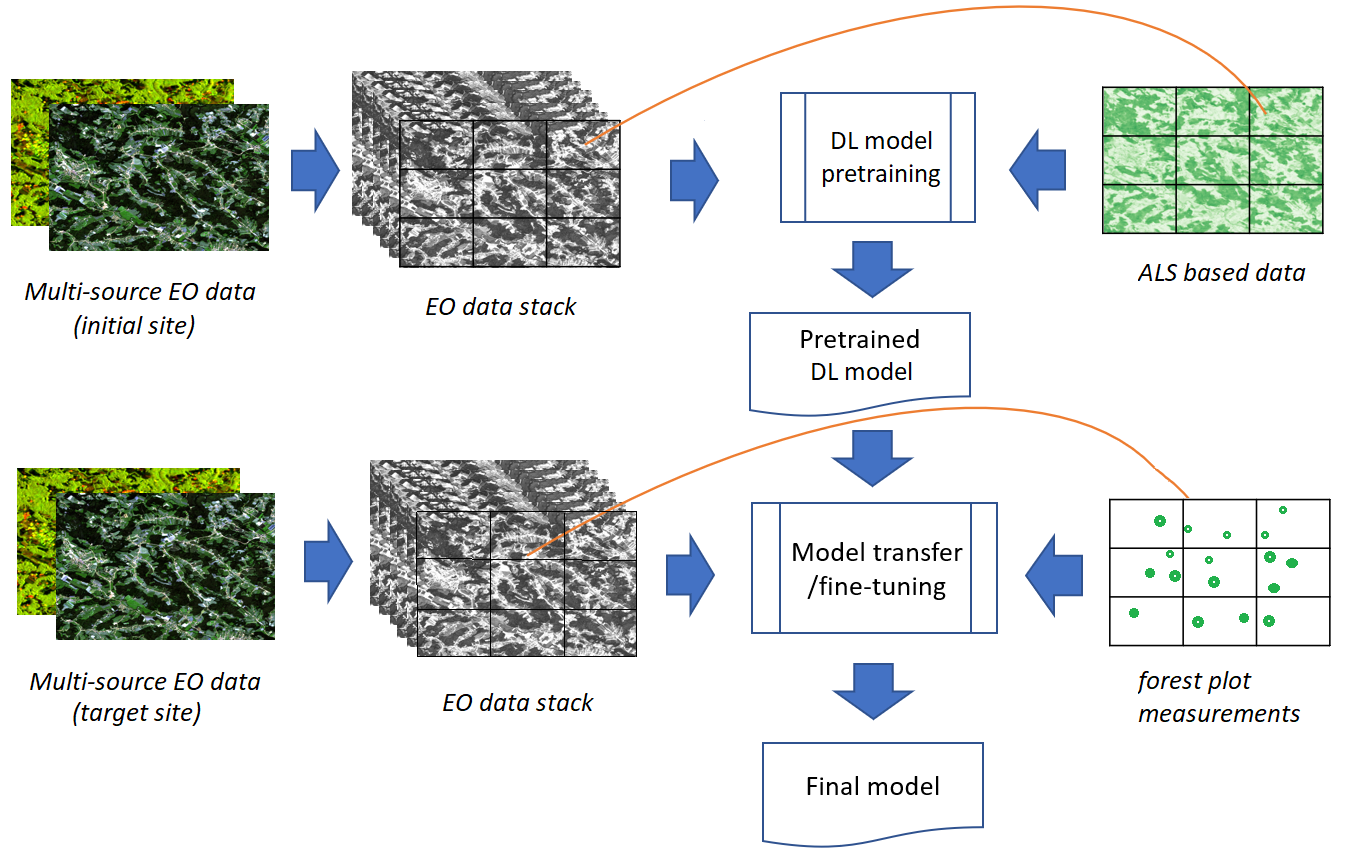}
\caption{Model transfer logic}
\label{fig: studylogic}
\end{figure*}

\begin{figure*}[htb]
\centering
\includegraphics[width=0.9\linewidth]{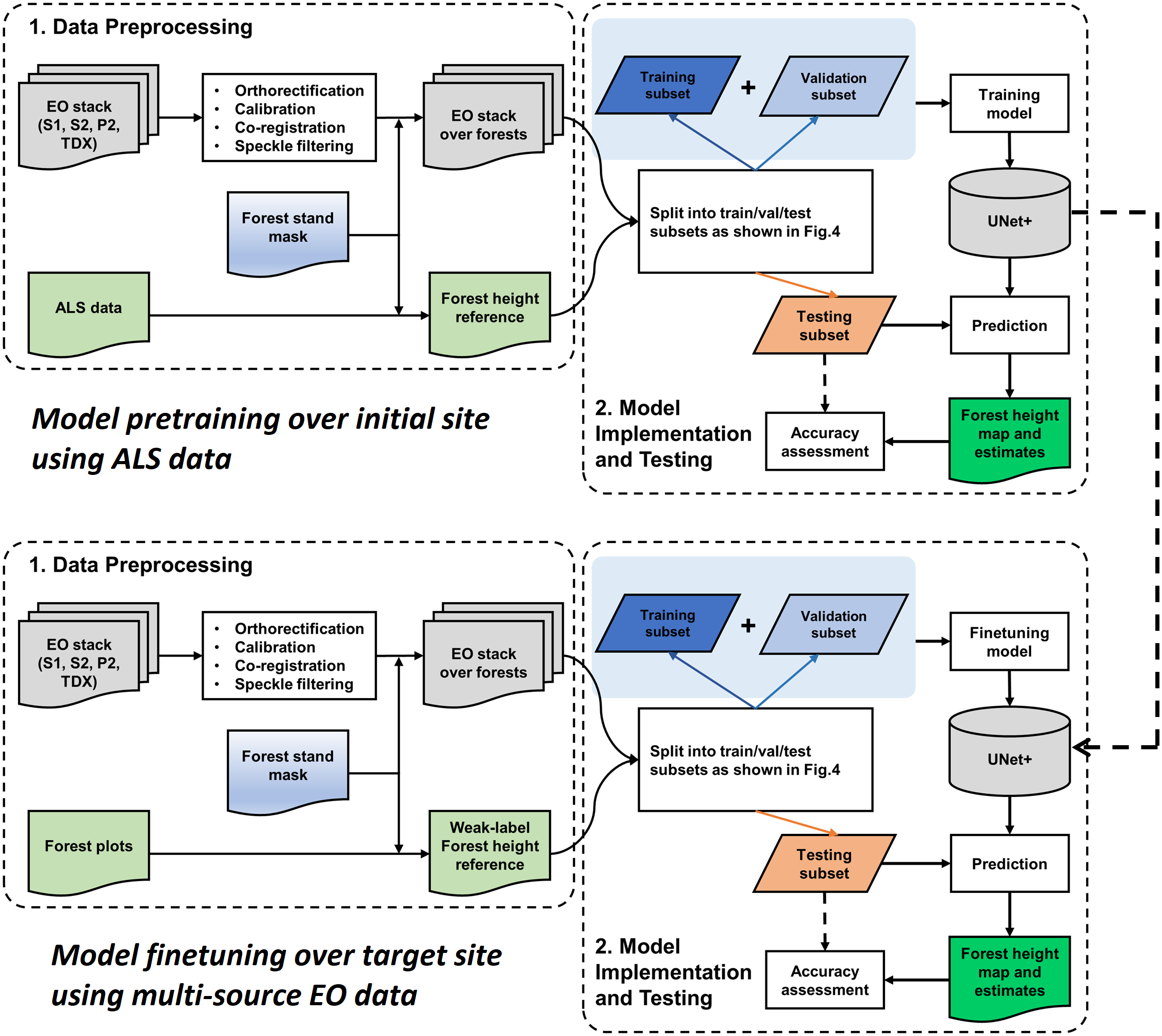}
\caption{Overall processing approach}
\label{fig: flow}
\end{figure*}

\subsection{Our approach}
Our suggested approach follows the study logic shown in Figure \ref{fig: studylogic}. Firstly, a SeUNet model is pretrained using multi-source EO-dataset and spatially explicit reference based on ALS data. This is done over the \textit{pre-training} site where such reference data are available. Both spectral and spatial forest signatures are learned as parameters of pre-trained UNet model. Various combinations of input satellite EO data are tested, resulting in a set of DL models.

In the second stage, the learned SeUNet parameters are used as the initial weights for \textit{model training over the target} site. In contrast to the pretraining site, only a very sparse reference dataset is available from the target site, represented by forest plots. The model is fine-tuned by including only pixels that have known reference value in the loss computation. Both the pretraining phase and DL model training over the target site are illustrated in Figure \ref{fig: flow}. We also investigate scenarios for which only a fraction (5-10\%) of originally available forest plots are used, and when several forest strata are underrepresented.

Lastly, our predictions are compared to predictions obtained using traditional EO-based forest inventory methods including multiple linear regression (MLR) and the popular k-Nearest Neighbours (k-NN) technique \cite{englhart2012,antropov2017,ge2021s1,antropov2022}. 
MLR is a basic regression approach often used for modeling the relationship between response variables such as GSV or forest biomass and SAR and optical image features \cite{berninger2018,rv2019,ge2020}. k-NN is an established non-parametric and distribution free method widely used for forest variable prediction \cite{tomppo2008}. Predictions are obtained as weighted linear combinations of attribute values in a set of k nearest units selected from a reference set of units with known values. The choice of these units is determined by a distance metric defined in the auxiliary variable space.

\subsection{Study sites, satellite SAR and optical data}
\begin{figure*}[htb]
\centering
\includegraphics[width=0.9\linewidth]{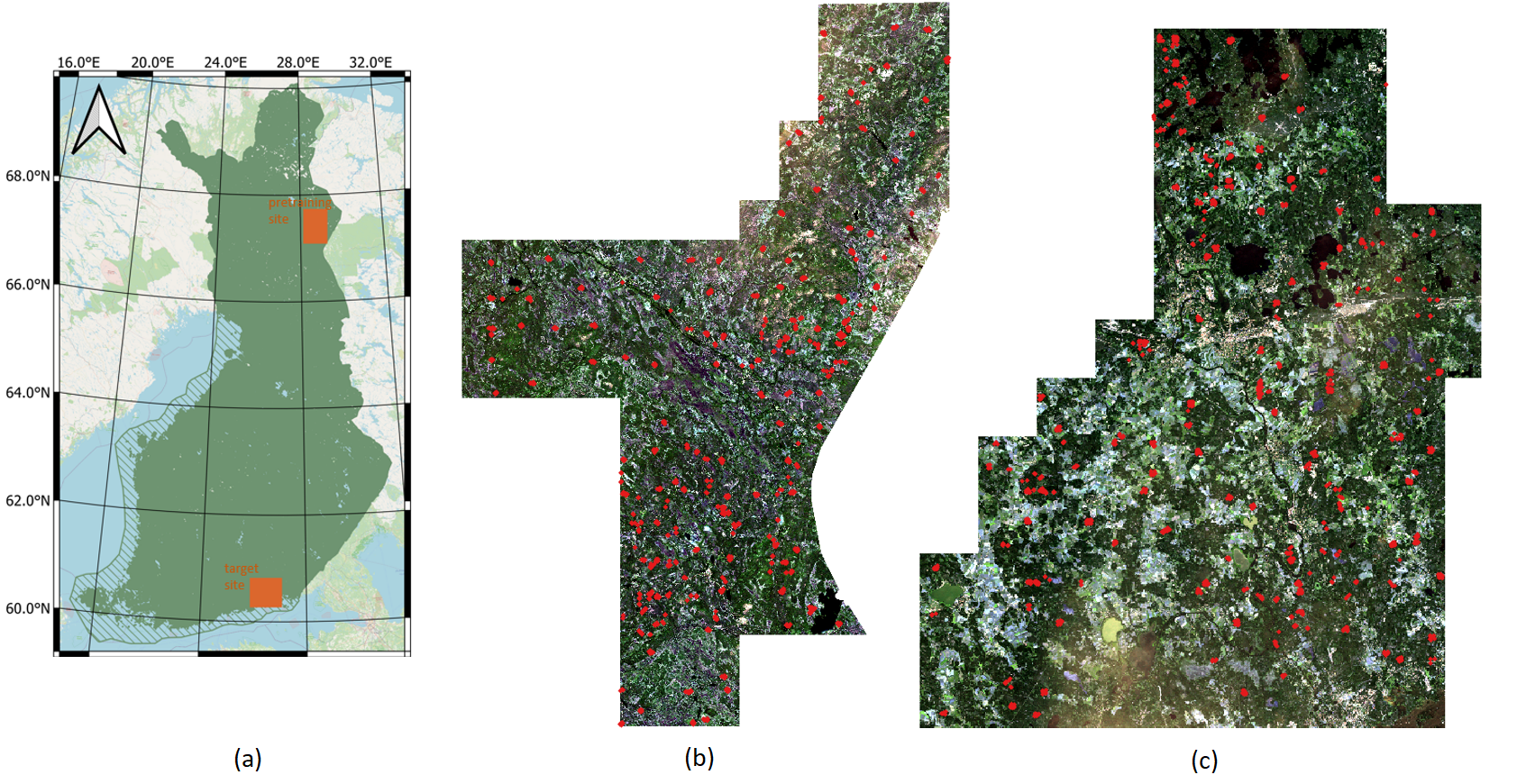}
\caption{Study site locations in Finland (a), Sentinel-2 natural color composite image over the Salla (pretraining) site (b), and over the Kouvola (target) site (c). Red dots designate forest plot locations.}
\label{fig: sites}
\end{figure*}

The study is performed in Finland over two geographically distinct areas, separated by around 700 km (Figure \ref{fig: sites}). Both sites feature boreal forests, one in the Northern Boreal zone and the other in the Southern boreal zone. The forests in the Northern Finland are in many aspects different from the forests in Southern Finland. This makes this pair of study sites particularly suitable for demonstrating the potential of DL model transfer.
The Northern site in Lapland, in the Salla municipality (hereafter called the Salla site) features a varying landscape of forests and open peatlands. The elevation above sea level ranges from less than 200 m to around 500 m. The forests are dominated by pine (~60\%) with some spruce (~20\%) and broadleaf forest (~20\%). The average volume in the study site is 95.3~m\textsuperscript{3}/ha, with an average height of 10.8 m. The forests in Lapland are typically sparser that in the south, with the average stem density around 1600 stems/ha in the Salla site. 
The Southern site between the towns of Kouvola and Kotka (hereafter called the Kouvola site) features typical Southeastern Finnish landscape with a mixture of forests and agricultural areas. The site has only minor elevation differences, with the highest points generally less than 100 m above sea level. The forest have around 40\% pine, more than 30\% spruce and the rest broadleaf trees. The average volume in the study site is 156.4~m\textsuperscript{3}/ha, with an average height of 15.0 m. Note that the volume and height are substantially greater than in the Salla site. Similarly, the stem density of around 2200 stems/ha in the Kouvola site is clearly greater than the stem density on the Salla site (1600 stems/ha)

\subsubsection{Satellite image data}
The EO dataset consists of a 14-channel tensor with Sentinel-2 image bands, Sentinel-1 yearly composite bands (VV and VH bands), as well as  ALOS-2 PALSAR-2 and TanDEM-X image features. The model training uses image patches with size 256 px $\times$ 256 px, all input images preprocessed and resampled to 10m $\times$ 10m pixel spacing.

\begin{itemize}
\item	Satellite optical data were represented by an ESA Sentinel-2 image acquired on July 12, 2018 over the Salla site and August 11, 2018 over the Kouvola site. The Level 2A surface reflectance product systematically generated by ESA and distributed in tiles of 110 × 110 km$^2$  was used. The Multi-Spectral Instrument on board Sentinel-2 satellites has 13 spectral
bands with 10 m (four bands), 20 m (six bands) and 60 m (three bands) spatial
resolutions. We used bands 2,3,4,5,8,11,12 in our analysis, that were found most useful for monitoring boreal forest in prior studies \cite{miettinen2021}.

\item	SAR data are represented by an annual composite of 39 Sentinel-1 images acquired during 2018 in the same geometry.
The original dual-polarization Sentinel-1A images available as GRD (ground range detected) products were radiometrically terrain-flattened and orthorectified with VTT in-house software using local digital elevation model available from National Land Survey of Finland \cite{rauste2007,hame2015igarss}. Final preprocessed images were in gamma-naught format \cite{small2011}.

\item	L-band SAR imagery was represented by JAXA mosaic produced from dual-pol ALOS-2 PALSAR-2 images acquired during 2018 were used. 

\item Interferometric SAR layers were represented by TanDEM-X images collected during summer 2018. TanDEM-X canopy height model is calculated via subtracting of TanDEM-X phase and topographic phase (calculated from local topographic map) in slant range followed by  phase-to-height conversion and geocoding obtained height product. It is later called interferometric canopy height model (ICHM) in the paper. Additionally, TanDEM-X coherence magnitude was used as an image feature layer. ESA SNAP software was used for calculating TanDEM-X image layers. 
\end{itemize}

\subsubsection{Reference data}
Over the initial pretraining site, ALS-based heights were used. ALS data were collected by National Land Survey of Finland during summer of 2018. Forest heights were estimatedfrom ALS point clouds as average elevation of forest classified points over ground layer within \SI{20}{}$\times$\SI{20}{m^2} pixel cells. In this way, a wall-to-wall coverage of the pretraining study site with the reference height information was obtained. 

Reference data over the target Kouvola site were represented by data for a sample of plots measured by the Finnish Forest Centre in 2018. The plots were circular with three different radii: 9 m in young and advanced managed forests with a relatively large tree density; 12.62 m in forest with a small stem density but usually large volume due to the mature development stage and 5.64 m in seedling stands. 
Altogether 1064 field plots were used. Two thirds of the plots, 709 of them, were used for model training (model transfer), while the remaining 355 (selected as every third plot after arranging the plots in the order by volume) were used for the uncertainty assessment.

\subsection{Implementation details}
In model pretraining, the wall-to-wall reference and EO data were first cropped into \SI{256}{px}$\times$\SI{256}{px} non-overlapping image patches in spatial dimension. The non-forested regions were removed by masking out corresponding areas on both EO data and reference data. Additionally, several patches with forest cover proportion less than 20\% were removed. 
In total, 614 image patches were prepared, half of which (307 patches) were randomly assigned to the testing subset, 10\% were assigned to the validation subset and the remaining patches were used in the model training. 
Later, after the data augmentations that included in-situ spatial shifting and rotations, 1433 augmented training patches were generated.

In model transfer, forest field plots were converted to rasters. We used the similar image patch cropping approach as described above, keeping only patches that have at least one plot-level reference within the patch. In total, there were 138 testing patches, 32 validation patches and 524 augmented training patches. We used the weak-labeled training and validation patches to fine-tune the pretrained model.

In both pretraining and fine-tuning processes, we used \textit{Adam} as the optimizer and \textit{OneCycleLR} as learning rate scheduler, the maximum learning rate was set to $10^{-2}$. We pretrained the model for 100 epochs in pretraining with weight decay factor of $10^{-4}$ to avoid overfitting.
The best checkpoint was determined based on the validation loss value and corresponding weights were saved for later fine-tuning the transferred model over the target site. In contrast to pretraining, we fine-tuned the transferred model for only 5 epochs.

For conventional methods, plot-level EO features were calculated using described datasets and data splits over the Kouvola target site and used in the model training.

The experiments were performed using Windows Server with Intel Xeon E5-2697 v4 CPU and NVIDIA RTX A5000 GPU accelerated by CUDA 11.7 toolkit. The SeUNet model was built with a neural network library, Pytorch 1.11.0. MLR and RF were implemented with Scikit-learn machine learning toolbox. 

\subsection{Accuracy metrics}
The prediction accuracy for the various regression models was calculated using the following accuracy metrics, {including root mean squared error (RMSE), relative root mean squared error (rRMSE), the~coefficient of determination (R$^2$) as follows: 
{
\begin{align}
\mathrm{RMSE}&=\sqrt{\frac{\sum_{i}\left(y_{i}-\hat{y}_{i}\right)^{2}}{n}}, \\
\mathrm{rRMSE}&=\frac{RMSE}{\bar{y}} \cdot 100 \%, \\
\mathrm{bias}&={\frac{\sum_{i}\left(y_{i}-\hat{y}_{i}\right)}{n}}, \\
\mathrm{R^{2}}&=1-\frac{S S_{\text {res}}}{S S_{\text {tot }}}=1-\frac{\sum_{i}\left(y_{i}-\hat{y}_{i}\right)^{2}}{\sum_{i}\left(y_{i}-\bar{y}\right)^{2}}, 
\end{align}}
where $y_i$ and $\hat{y}_{i}$  are reference and predicted values of forest height for $i$-th plot, 
$\bar{y}$ is the mean value of all plots and $n$ is the total number of plots.

\section{Results}

\subsection{Prediction Performance over Pretraining (Salla) Site}

\begin{figure*}[htb]
\centering
\includegraphics[width=\linewidth]{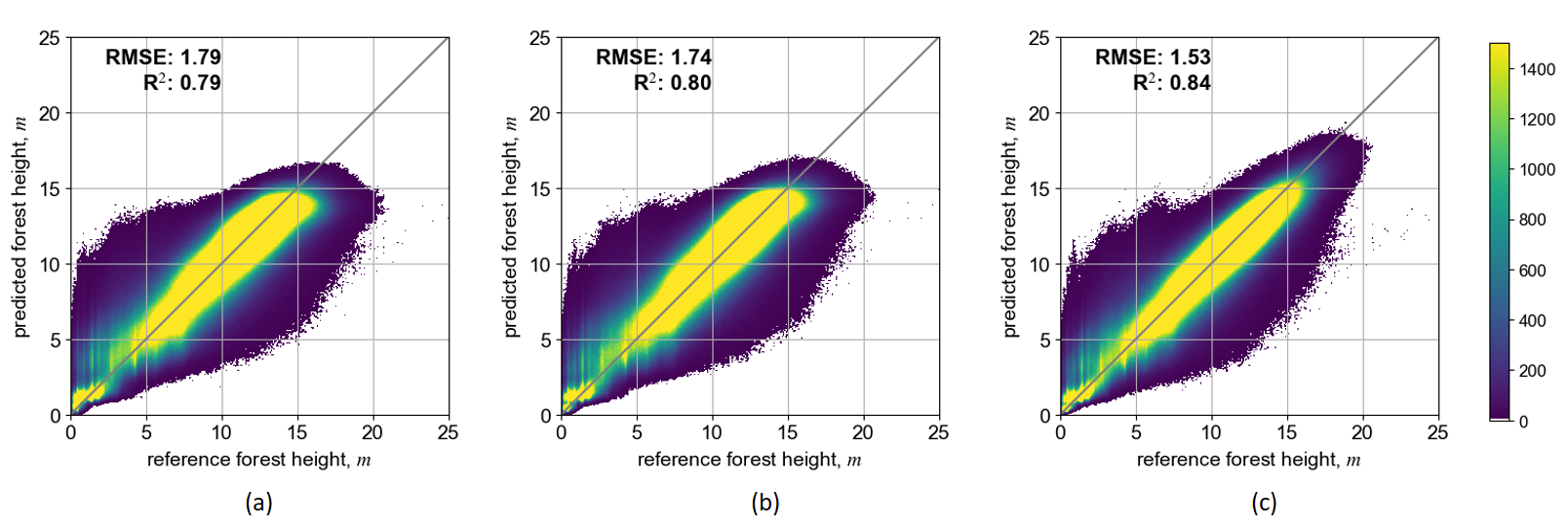}
\caption{Scatterplots illustrating prediction performance of pretrained SeUNet model using ALS reference data: a) \textit{S2-Lapland} model, b) \textit{S1S2-Lapland} model, c) \textit{MS-Lapland} model}
\label{fig: als-scatter}
\end{figure*}

Figure \ref{fig: als-scatter} shows scatterplots after the model pretraining on original/initial dataset over the Salla site in Lapland. Three combinations of input EO datasets resulted in three distinct DL models: 1) \textit{S2-Lapland}, with only Sentinel-2 data; 2) \textit{S1S2-Lapland}, where both Sentinel-2 and Sentinel-1 data were included, and 3) \textit{MS-Lapland}, where multi-source data included additionally ALOS-2 PALSAR-2 and TanDEM-X data layers. 

Depending on the input EO dataset, prediction accuracy of the SeUNet model (calculated on testing image patches not involved in the training process) varied. Scenario that included only Sentinel-2 bands had somewhat smaller prediction accuracy (RMSE=1.79 $m$, rRMSE=18.8\%,  R$^2$=0.79) while adding Sentinel-1 layer slightly improved the prediction performance (RMSE=1.74 $m$, rRMSE=18.2\%, R$^2$=0.80). Importantly, these results can be achieved using freely available Copernicus datasets.  When adding ALOS-2 PALSAR-2 mosaic (with well-known L-band sensitivity to forest growing stock volume) or TanDEM-X data (with well-known sensitivity to vertical forest structure) the prediction accuracies increased to RMSE=1.53 $m$, rRMSE=16.1\% and R$^2$=0.84. Importantly, those predictions are done at 10 $m$ pixel resolution, and accuracy estimates increase when aggregating to coarser mapping units.

\subsection{Prediction Performance over Target (Kouvola) Site}
Results of "blindly" applying pretrained models (without fine-tuning with in-situ forest plots) over the target Kouvola site are shown in the upper row of Figure \ref{fig: scatter_seunet}. Prediction performance is not satisfactory, with RMSE in the range 5-7 $m$ (40-50\% rRMSE), strong negative systematic prediction error of 2.3-2.8 $m$ present in all model predictions, and apparent signal-saturation effects clearly visible for taller trees. Such performance can be attributed to differences in both forest structure and EO images (spectral, calibration, seasonal changes). Similar albeit smaller effects could be expected if the target site was the same as pretraining, and only EO images acquired at another time were used in the prediction (i.e., at least the forest structure would be the same). Prediction performance was problematic for all non-finetuned DL models and EO data inputs, with slightly more accurate predictions for the \textit{MS-Lapland} model.

After the SeUNet model fine-tuning, accuracies strongly increased. Scatterplots for corresponding models are shown in Figure \ref{fig: scatter_seunet}, bottom row.  Detailed results are collected in Table \ref{tab: target_stats} for all models as well as for the benchmark MLR and kNN methods. Corresponding scatterplots for the benchmark models are shown in Figure \ref{fig: scatter_baseline}.

Using additional explanatory EO variables improved prediction accuracy in all cases, for both baseline methods and for developed models. The multi-source dataset demonstrated the most accurate predictions for all methods including benchmark methods, similar to model pretraining with ALS data.

Achieved prediction accuracies are considerably improved compared to applying non-finetuned methods and compared to traditional machine learning approaches.  

\begin{figure*}[htb]
\centering
\includegraphics[width=0.9\linewidth]{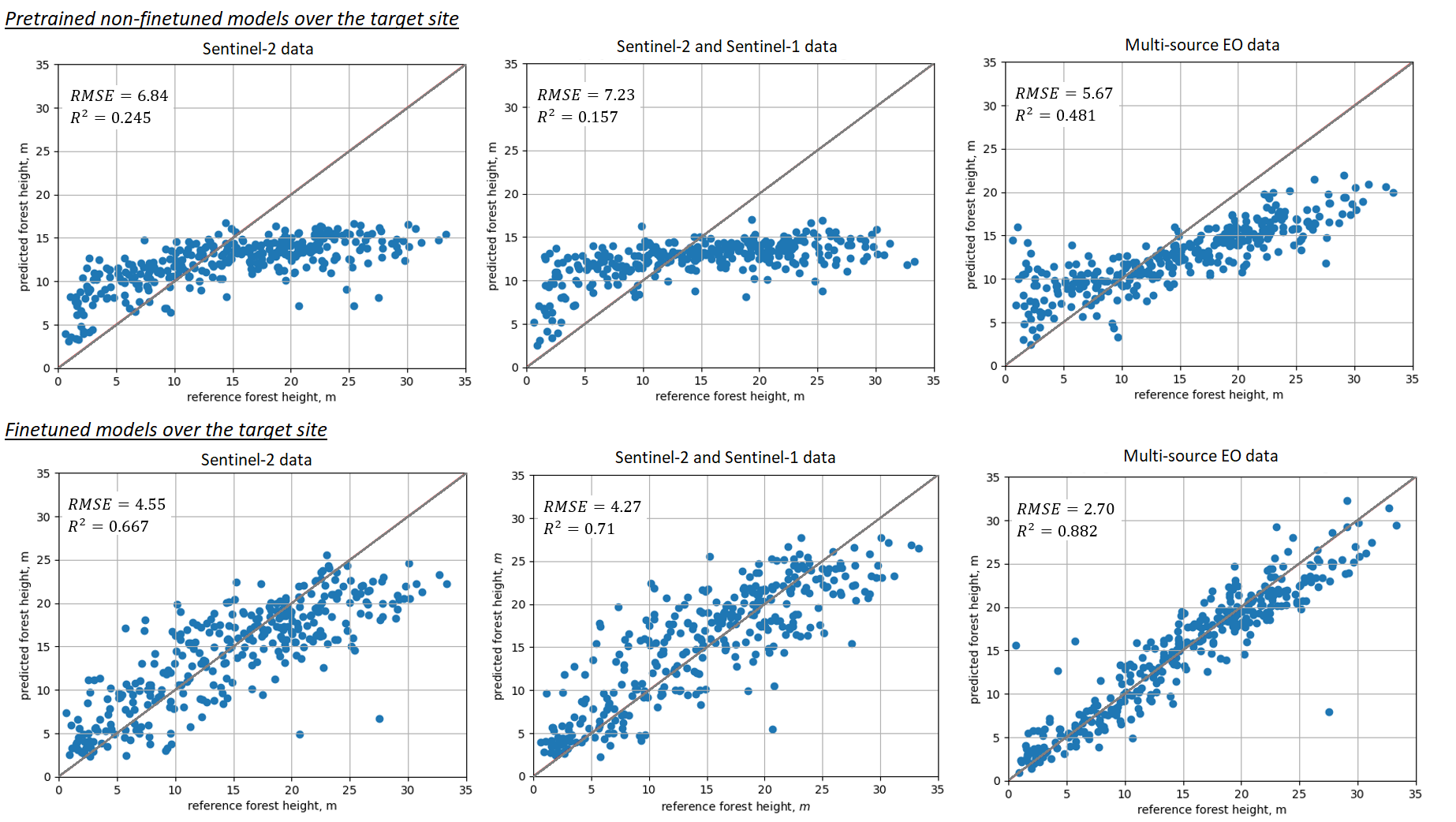}
\caption{Scatterplots illustrating prediction performance for SeUNet model: {upper row - use of pretrained non-finetuned models}; bottom row - using fine-tuned models;  1st column - Sentinel-2 data; 2nd column - combined Sentinel-2 and Sentinel-1 images, 3rd column  - all available images (Sentinel-2, Sentinel-1, ALOS-2 PALSAR-2, TanDEM-X)}
\label{fig: scatter_seunet}
\end{figure*}

\begin{figure*}[htb]
\centering
\includegraphics[width=0.9\linewidth]{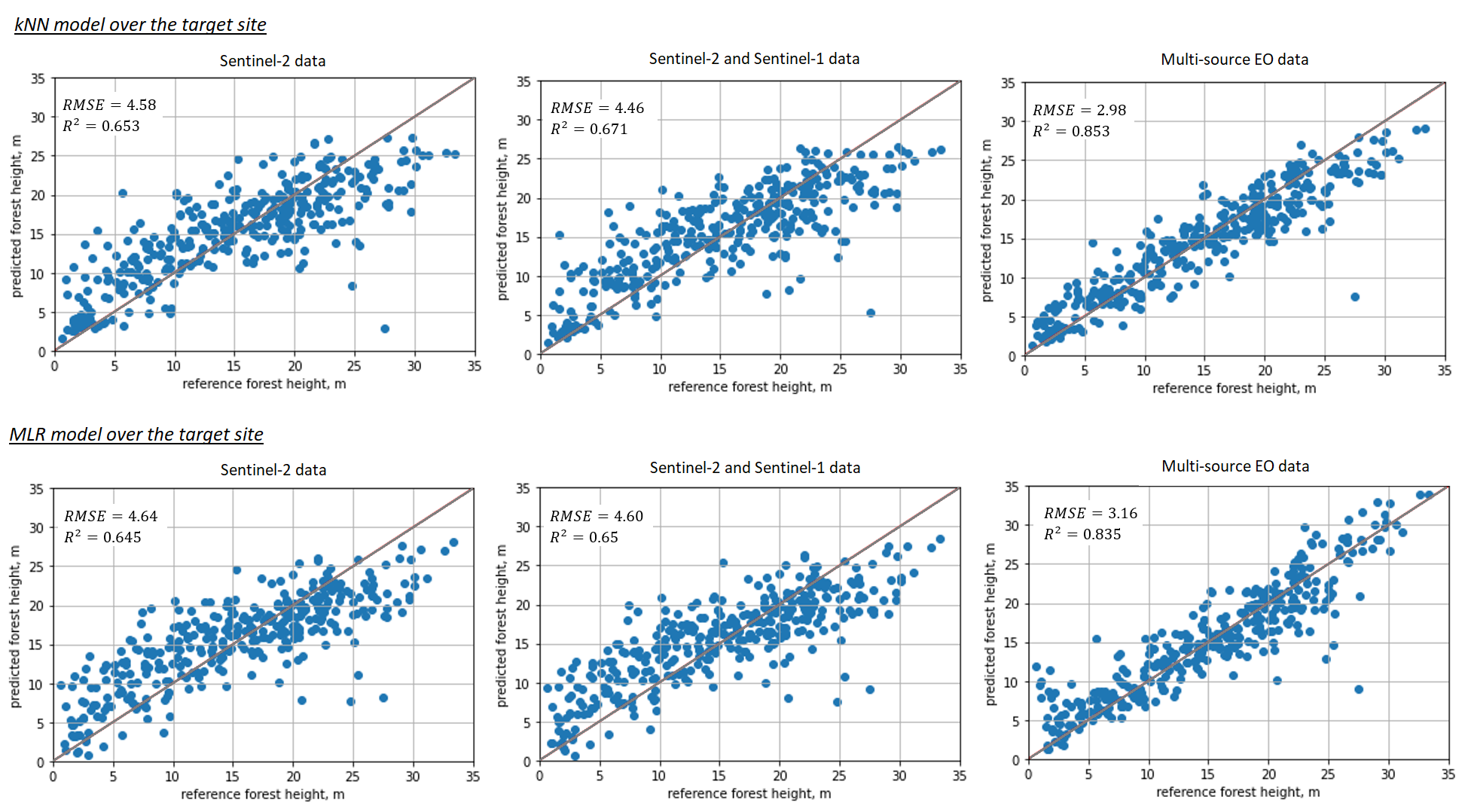}
\caption{Scatterplots illustrating prediction performance for baseline methods: upper row - kNN, bottom row - MLR; 1st column - Sentinel-2 data; 2nd column - combined Sentinel-2 and Sentinel-1 images, 3rd column  - all available images (Sentinel-2, Sentinel-1, ALOS-2 PALSAR-2, TanDEM-X)}
\label{fig: scatter_baseline}
\end{figure*}

\begin{table}[htb]
\centering
\caption{Prediction accuracy statistics for Kouvola (target) site}
\label{tab:results_px}
\begin{tabular}{@{}lcccc@{}}
\toprule
                  & \textbf{RMSE, m} & \textbf{RMSE\%} & \textbf{Bias} & \textbf{R$^2$}  \\
\multicolumn{5}{l}{\textit{Sentinel-2 data}}                                    \\ \midrule
\textbf{MLR}      & 4.64    & 30.6      & 0.30    & 0.645   \\
\textbf{kNN}     & 4.58    & 30.3       & 0.28    & 0.650   \\
\textbf{SeUNet, non-finetuned}     & 6.84    & 45.8       & -2.75    & 0.245   \\
\textbf{SeUNet, finetuned}   & 4.55    & 30.3       & -0.77   & 0.667   \\
\midrule
\multicolumn{5}{l}{\textit{Sentinel-2 and Sentinel-1 data}}                                    \\ \midrule
\textbf{MLR}      & 4.60    & 30.4      & 0.30    & 0.65   \\
\textbf{kNN}     & 4.46    & 29.5       & 0.26    & 0.67   \\
\textbf{SeUNet, non-finetuned}     & 7.23    & 48.4       & -2.48    & 0.157   \\
\textbf{SeUNet, finetuned}   & 4.27    & 28.6       & 0.53   & 0.71   \\
\midrule
\multicolumn{5}{l}{\textit{Multi-source EO data}}                                 \\ \midrule
\textbf{MLR}      & 3.16    & 20.9      & 0.23    & 0.835   \\
\textbf{kNN}     & 2.98    & 19.7       & -0.40    & 0.853   \\
\textbf{SeUNet, non-finetuned}     & 5.67    & 38.0       & -2.32    & 0.481   \\
\textbf{SeUNet, finetuned}   & 2.70    & 18.1       & -0.25   & 0.882   \\
 \bottomrule
\end{tabular}
\label{tab: target_stats}
\end{table}

\subsection{Model stability with scarce or missing reference data in the Kouvola site}
Additionally, we checked the stability/resilience of the suggested model and baseline approaches with respect to scarce (only 5-10\% of all plots are available over target site, suitable for SeUNet model transfer or baseline traditional model training) or completely missing data (e.g., forest plots with tall trees or forest plots with short trees are completely missing from the target site). The baseline method for comparison was kNN, which is widely used in forest inventory mapping\cite{mcrob2007}, and which also demonstrated superior performance compared to MLR also in our experiments. Using multi-source EO data, specific considered cases that simulate realistic scenario of sparse measurements or lack of specific forest strata over target site were:
\begin{itemize}
  \item only 35 plots (5\% of original training sample) are used in model training (model transfer)
  \item small-biomass plots with forest heights less than 10 meters are completely removed from the training dataset.
  \item tall forest plots with heights exceeding 25m are completely removed from the training dataset.
\end{itemize}

It can be seen that in all the "extreme" cases, illustrated in Figure \ref{fig: scatter_extreme}, classical nonparametric approaches such as kNN start to fail when reference data are scarce or non-representative (e.g., missing the smaller or larger end of the height range). The difference in R$^2$ and RMSE is substantial between fine-tuned SeUNet and kNN methods, but most importantly the range of predicted heights is distorted, and predictions can not understandably reach certain (small or large) values that fall out of the range of forest plot measurements. For example, when young forest plots are missing in the training set (Figure 8, 2nd column), kNN fails to produce any estimates less than 10 m. The same effect can be observed for forest plots with tall trees, with the tallest predictions of $\sim$25 m for kNN, while the SeUNet predictions reach 30 m. When a small number of plots is used, kNN predictions start to approach the average height. The effect would become even more pronounced if the number of plots is decreased further, e.g. to 10 reference plots. The model transfer is still expected to work in that situation, provided that the reference dataset for the target site includes sample units with short and tall forests.

\begin{figure}[htb]
\centering
\includegraphics[width=1.05\linewidth]{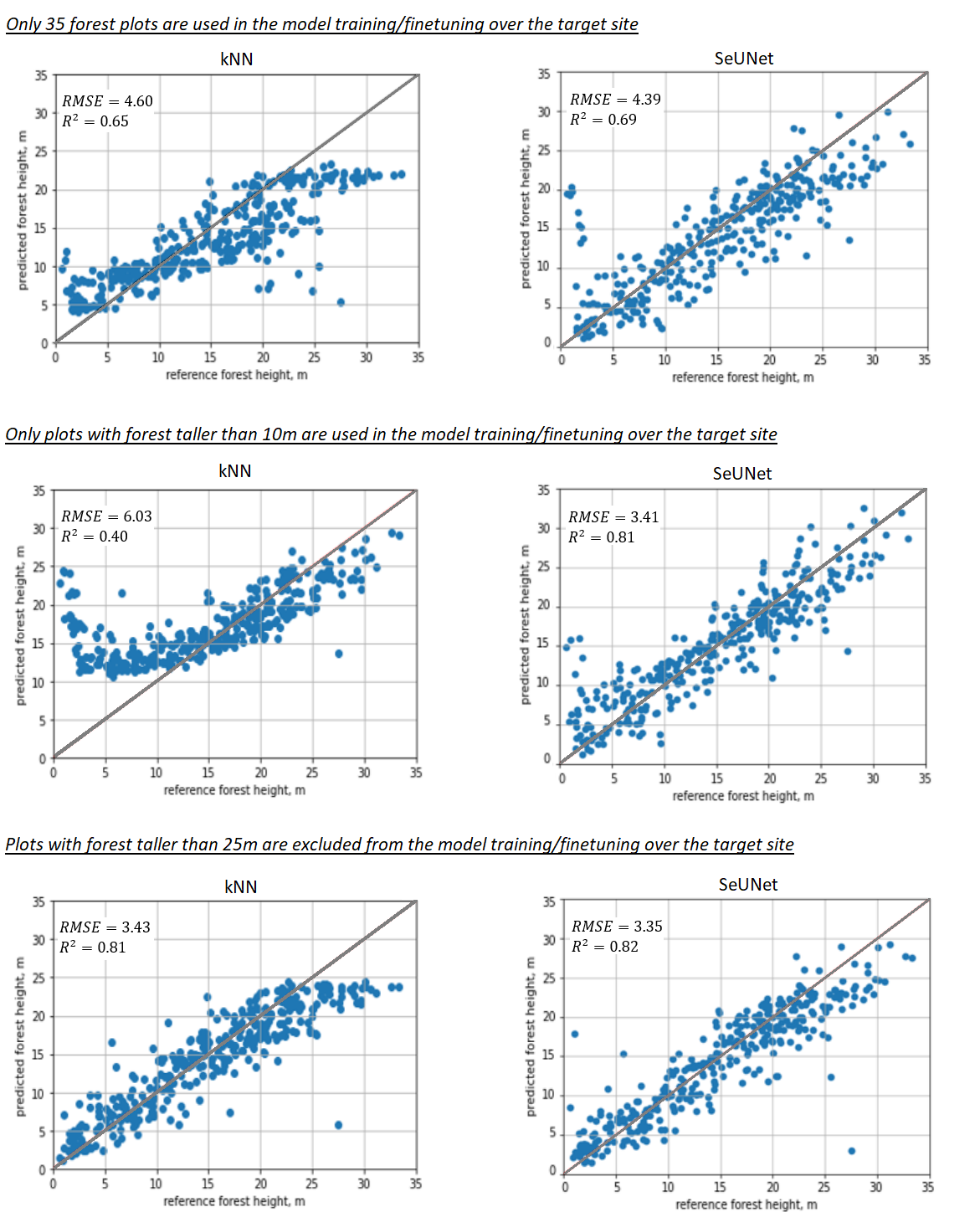}
\caption{Scatterplots illustrating prediction performance for nonparameteric models: 1st column - kNN, 2nd column - SeUNet, 1st row - scarce training sample (35 plots used), 2nd row - plots smaller than 10m were removed during model training/finetuning in the target site, 3rd row - plots with forest taller than 25 meters were absent in model training/finetuning over the target site}
\label{fig: scatter_extreme}
\end{figure}

\begin{figure*}[htb]
\centering
\includegraphics[width=0.9\linewidth]{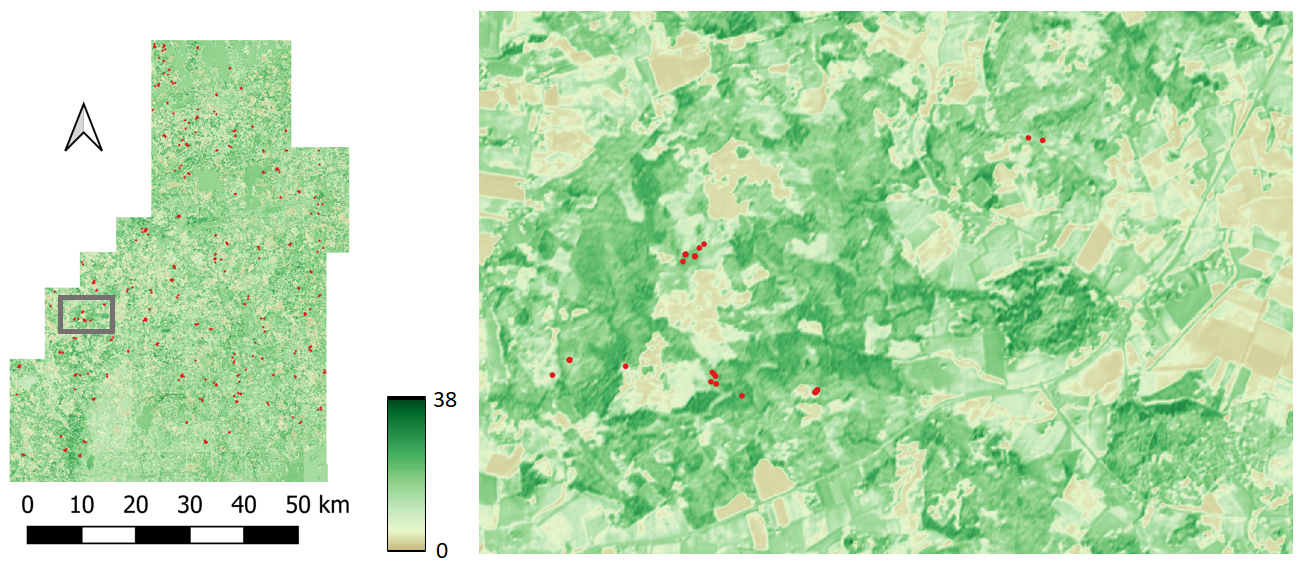}
\caption{Predicted forest height map over the Kouvola (target) site and zoomed in fragment. SeUNet model has been used in the prediction. }
\label{fig: predict_map}
\end{figure*}

\section{Discussion}

\subsection{Overall discussion on performance across various models and EO datasets}

Figure \ref{fig: scatter_seunet} clearly demonstrates increased accuracy for SeUNet model prediction when using transfer learning mechanism. As visible from scatterplots in the upper row, the prediction without transfer learning does not follow the 1:1 line very well, showing strong systematic negative prediction error and saturation effects. The heights of taller forests greater than \SI{15}{m} are underestimated, while height predictions of forests shorter than \SI{10}{m} tend to be overestimated. In the best case when multisource satellite data are used, the mean prediction error exceeds \SI{2}{m}. 

Within the transfer learning, the finetuning process compensates  for the disadvantage of insufficient or erroneous (because of different forest structure and EO data) representation information. Representation features themselves, particularly spatial context learned from source site are transferred to the target site.
As shown in the bottom row of Figure \ref{fig: scatter_seunet}, when using multisource EO data, the rRMSE is reduced by 20\% compared to the scenario for which the non-finetuned model is applied, and the mean prediction error is accordingly reduced to less than \SI{0.3}{m}.

Additionally, we compare the performance of SeUNet to other traditional machine learning methods, such as kNN and MLR. The deep learning model generated more precise forest height predictions, especially when multi-satellite data are used. Achieved RMSE=\SI{2.70}{m} with rRMSE=18.07\% is several percentage units smaller than obtained with kNN and MLR. From the scatterplots, we can clearly observe a closer linear relationship between predictions and reference. The improvement can be attributed to learned spatial representations from pre-training site. This is in line with common observation that use of textural information can help improve predictions of forest variables \cite{schmullius2015}.

Regarding the role of different EO datasets, combining SAR and optical datasets provided more accurate predictions than use of optical or SAR datasets alone, even though gain from adding Sentinel-1 to Sentinel-2 was limited compared to using Sentinel-2 data only. In Figure \ref{fig: scatter_seunet} and Figure \ref{fig: scatter_baseline}, the bias is slightly reduced when adding Sentinel-1 data. One possible reason of limited improvement is that Sentinel-1 data are not the optimal dataset for forest height estimation. In contrast, adding L-band ALOS-2 PALSAR-2 and especially interferometric TanDEM-X image layers strongly increased the accuracies; such effects were observed over both the pretraining site with ALS data, and over the target site. With multi-source data, considerable reduction in rRMSE was observed not only for SeUNet but also for kNN and MLR, which can be explained by high sensitivity of TanDEM-X to vertical structure of forests \cite{kugler2015,olsk2016}.

\subsection{Comparison with Prior Studies}

To date, numerous methods and remotely sensed data combinations have been used for forest height estimation in boreal and temperate forests \cite{huang2022,luo2023,zhang2022,astola2019,miettinen2021}. Reported accuracies for boreal forest height mapping range typically in the order of 30-40\% rRMSE in these studies. The results achieved in the present study can therefore be considered quite comparable to earlier methods, although in this study a model transfer between two highly different study sites was performed. 

In the boreal region, reported forest height accuracies with Sentinel-2 and Landsat data have been 35–60\% rRMSE at forest plot level\cite{astola2019,astola2021,miettinen2021}, while the proposed model transfer approach reached 30\% rRMSE at plot level with Sentinel-2 data only, and 18\% rRMSE with multi-source EO dataset.
 Predictions obtained with ML models and Sentinel-2 data are within the same accuracy range as in recent published studies using Sentinel-2 and Landsat \cite{astola2019}, while our predictions using DL models are more accurate for similar EO data combinations. The literature on using SAR data for forest height prediction is limited, with most studies conducted at forest-stand level, but our tree height predictions are at the same accuracy level or even greater than reported for retrievals with TanDEM-X interferometric SAR data that are considered very suitable for vertical forest structure retrieval\cite{praks2012,kugler2015,olsk2016}.
 Regarding coupling of ALS data with satellite EO data in our pretraining in Lapland, our results are in line with other similar studies \cite{LI2020102163}. Use of recurrent and fully convolutional neural networks with fully segmented labels and Sentinel-1 time series or combined SAR and optical data provided accuracies on the order of 17-30\% rRMSE that are similar to results in our work over pretraining site in Lapland \cite{ge2022improved,ge2022lstm,lang2019country,BECKER2023269}. 
 Inversion of TanDEM-X images acquired over  Estonian hemiboreal and Canadian boreal forests provided accuracies with RMSE in range of 3–4 m and correlation coefficients R$^2$ larger than 0.5 \cite{chen2016, olesk2015,praks2018, praks2012,chen2018}. Such results were achieved using various sets of TanDEM-X data and simplified semi-empirical parametric models. Other reports indicate typical error levels around 4.8 m RMSE for TanDEM-X datasets in comprehensive studies using those models \cite{SCHLUND2019101904}.

 Our results with \textit{MS-Lapland} SeUNet model fine-tuned using forest plots in Kouvola are more precise thus indicating useful synergy of SAR and optical datasets in boreal forest parameter mapping and thus benefits of using combined multi-source datasets.

\subsection{Outlook}

The first demonstration of the model transfer of a deep learning model for forest variable estimation performed in this study indicates substantial potential of such approaches for operational forest monitoring tasks. 
Demonstrated study can be considered a "geographic transfer", while also "temporal transfer" is possible when a model is pretrained on one year's EO data, and then later used over the same site but with EO data collected in another year. Obviously, a combination of those is possible as well. 
 Both types of model transfer hold great potential for forest monitoring.

The geographic model transfer, such as the one used in this study, extends the utility of deep learning models into areas that do not have suitable training data. Knowing the sporadic availability of ALS data over boreal and temperate forests, this would greatly expand the usability of deep learning models. It must be remembered that in this study the transfer was performed within the same boreal zone, albeit between two forest areas with different structural characteristics. Further studies are needed to investigate the limitations of transferability of models between different biomes with greater structural differences (e.g. in species composition). Similarly, transferability of models between different remotely sensed data combinations would be required. This would further broaden the benefit of model transfer, by allowing flexible use of the best possible dataset combinations in the target area, regardless of the dataset used in the model development.

Temporal model transfers, which were not tested in this study at all, would improve the efficiency of high frequency forest monitoring even in areas that have ALS data coverage for some years. 
With a limited number of field plots measured in the following (or preceding) years, availability of pretrained DL model and model transfer technique would allow use of the fine-tuned model in the years that do not have any ALS data available from the area. This would improve the frequency and consistency of forest monitoring in the area, which is something that is in high demand as the monitoring and reporting requirements for forest owners are constantly increasing.

The results of this study also indicate also that model transfer can be performed with sub-optimal non-representative field plot reference data (e.g., with very few plots or limited range of observations). Limitations of existing field datasets can be overcome by transferring a model that is trained with optimal datasets into the target area (or year) by fine-tuning it with existing sub-optimal field plots. Alternatively, limited field campaigns can be conducted to collect sufficient data for fine-tuning. This increases the efficiency and reduces the costs of forest inventories enabling increase in temporal frequency in an economical manner. Thus, the approach presented in this study has the potential to support forest owners in meeting the increasing forest monitoring and reporting requirements.

In our opinion, further improvement from an operational viewpoint, and also in terms of prediction accuracy, can be gained by improving "initial" pretrained deep learning model that is further fine-tuned. The Lapland model in this regard was limited as did not feature the whole range of forest height and biomass values. Reference data from the whole country, such as Finland, can be used to establish such baseline deep learning model for taiga forests. Our further work will focus on scaling the demonstrated approach and establishing such baseline multi-source EO models for various forest biomes. Another direction is incorporating other UNet+ models, as well as other semi-supervised  convolutional and recurrent models to form an extended set of deep learning modeling approaches.

\section{Conclusions}

This is the first demonstration of deep learning model transfer in the context of EO based forest inventory using multi-source optical and SAR data, for which ALS data were used in the model pretraining, and only a limited sample of forest plots was used in the target area to fine-tune the model. This approach facilitated production of more accurate predictions compared to more traditional modeling and machine learning approaches, particularly when reference data were incomplete, very sparse or underrepresented. 

The proposed approach offers new perspectives in multi-source EO based forest mapping using pretrained deep learning models and a sparse set of forest plots. We demonstrated that such an approach can deliver greater accuracies compared to traditional machine learning methods, and importantly it is also quite robust to underrepresented or scarce forest plot data that are used in fine-tuning - when other machine learning models completely fail. This opens new perspectives in operational forest management and producing timely and updated forest inventories using EO dataset.

\section*{Acknowledgments}
This research was funded by European Space Agency within project on Forest Carbon Monitoring, grant agreement number 4000135015/21/I-NB. The processing costs were covered by the ESA Network of Resources (NoR) initiative. S.G. was supported by the National Natural Science Foundation of China under Grant 62001229, Grant 62101264 and Grant 62101260.

\bibliographystyle{IEEEtran}
\bibliography{main.bib}

\vfill

\end{document}